\documentclass[11pt,a4paper,showpacs]{article}
\usepackage{moriond,epsfig,slashed}

\def\mco{\multicolumn}

\def\be{\begin{equation}}
\def\ee{\end{equation}}
\def\bea{\begin{eqnarray}}
\def\eea{\end{eqnarray}}

\newcommand{\mt}{\ensuremath{M_T}}
\newcommand{\pte}{\ensuremath{p_T(e)}}
\newcommand{\ptmu}{\ensuremath{p_T(\mu)}}
\newcommand{\mete}{\ensuremath{\slashed{E}_T(e)}}
\newcommand{\metmu}{\ensuremath{\slashed{E}_T(\mu)}}
\begin{document}
\vspace*{1.5cm}
\begin{flushright}
FERMILAB-CONF-12-103-E
\end{flushright}
\vspace*{1.5cm}
\title{Precise measurements of the $W$ mass at the Tevatron\\and indirect constraints on the Higgs mass}

\author{Rafael C. Lopes de S\'a\\ (on behalf of the D\O\ and CDF collaborations)}
\address{Department of Physics and Astronomy, Stony Brook University\\Stony Brook, New York 11794, USA\\ {\rm rafael.lopesdesa@stonybrook.edu}
}

\maketitle\abstracts{
I describe the latest D\O\ and CDF $W$ boson mass measurements. The D\O\ measurement is performed with $4.3\, fb^{-1}$ of integrated luminosity in the electron decay channel with a data set of $1.68\times 10^{6}$ $W$ candidates. The value of the $W$ boson mass measured by D\O\ is $M_W = 80.375\pm 0.023\, GeV$ when combined with the previously analyzed $1\, fb^{-1}$ of integrated luminosity. The CDF measurement uses $2.2\, fb^{-1}$ of integrated luminosity in both electron and muon decay channels with a total of $1.1\times 10^{6}$ $W$ candidates. The value of the $W$ boson mass measured by CDF is $M_W = 80.387\pm 0.019\, GeV$. I report the combination of these two measurements with previous Tevatron measurements and with the LEP measurements of the $W$ boson mass. The new world average is $M_W = 80.385\pm 0.015\, GeV$. I discuss the implications of the new measurement to the indirect measurement of the Standard Model Higgs boson mass.\\\ \\\
}

\section{Introduction}\label{sec:Intro}

The electroweak sector of the Standard Model is described by a $SU(2)\times U(1)$ gauge theory with symmetry spontaneously broken by a Higgs doublet to account for the observed mass of the gauge bosons. All particles in the spectrum of this theory have been experimentally observed but the physical Higgs boson. Direct searches limit, at $95\%\, C.L.$, the possible values of the Higgs boson mass to the low-mass range of $115 \mbox{ -- } 127\, GeV$ or above $600\, GeV$~\cite{NewHiggs}.

Due to gauge structure of the symmetry-broken theory, the value of masses and coupling constants are not independent. The Standard Model prediction for the value of the $W$ boson mass has been calculated to full two-loop order~\cite{Awramik:2003rn} and, due to their large masses, is strongly dependent on the values of the $Z$ boson, Higgs boson and top quark masses. This prediction together with other electroweak observables can be used to indirectly measure any electroweak parameter and, in particular, the yet to be measured value of the Higgs boson mass~\cite{ALEPH:2005ab}.

Among all observables in global electroweak fits, the $W$ boson and top quark masses play the most important roles, since their direct and indirect measurements have similar uncertainties and, therefore, small improvements in either have a strong impact on indirect constraints of the Higgs boson mass and on the determination of the overall consistency of the electroweak sector of the Standard Model.

Precision measurements of the $W$ boson mass were performed by all LEP experiments and by both D\O\ and CDF experiments at the Tevatron collider~\cite{OldW}. The Tevatron was a $p\bar{p}$ collider working at $1.96\, TeV$ of center of mass energy. The impossibility of fully reconstructing the final state with the undetected neutrino is a major challenge that has to be dealt with when measuring the $W$ boson mass in a hadron collider. However, due to the large number of events recorded, both CDF and D\O\ are now able to measure the $W$ boson mass more precisely than the final LEP combined result. The world average before the results presented in this note was $80.401\pm 0.023\, GeV$.

\section{Measurement strategy}

As discussed in Sec.~\ref{sec:Intro}, the main feature of measuring the $W$ boson mass in a hadron collider is the impossibility of knowing the initial longitudinal momentum of the parton collision. This not only implies that the uncertainty in the measured value of the $W$ boson mass will have a large contribution from the uncertainties in the parton distribution functions inside the proton, but also that the measured phase-space of the $W$ leptonic decay is always incomplete because it is impossible to determine the neutrino longitudinal momentum.

Both D\O\ and CDF measurements explore the measured lepton and neutrino transverse momenta to determine the $W$ boson mass. Binned maximum-likelihood fits to transverse kinematical distributions are used to extract the value of the $W$ boson mass and its uncertainty. D\O\ uses both the electron transverse momentum and the transverse mass distributions. The transverse mass is defined as:

\begin{equation}
M_T(e,\nu) = \sqrt{2\left[p_T(e)\slashed{E}_T(e) - \vec{p}_T(e)\cdot\vec{\slashed{E}}_T(e)\right]}
\end{equation}
where $\vec{\slashed{E}}_T(e)$ is the missing transverse momentum of the event.

The CDF measurement uses six different distributions to extract the $W$ boson mass: the lepton and neutrino transverse momenta distribution and the transverse mass distribution in both electron and muon decay channels.

The different observables are not fully correlated since their measured distributions are shaped by different effects. Transverse momenta distributions are heavily shaped by the $W$ boson transverse momentum and, therefore, are sensitive to details of the initial state radiation that needs to be carefully modeled. The transverse mass distribution, on the other hand, is less sensitive to the $W$ boson transverse momentum but is shaped by detector resolution effects.

With the increasing experimental precision of the measurements, the more systematically limited extraction of the $W$ boson mass using the neutrino transverse momentum distribution becomes irrelevant in the final combination and D\O\ does not use this measurement, although it was performed and shown to be statistically consistent with the two others.

\section{Event selection}

In their measurement, CDF analyzes $2.2\, fb^{-1}$ of integrated luminosity. Events are required to have a single central ($|\eta| < 1$) muon or electron with transverse momentum in the range $30 < p_T(\ell) < 55\, GeV$. The neutrino transverse momentum is required to be in the same range $30 < \slashed{E}_T(\ell) < 55\, GeV$ and the pair transverse mass in $60 < M_T(\ell,\nu) < 100\, GeV$. Events with large $W$ transverse momentum, when the mass information is too diluted, are suppressed by requiring that the hadronic recoil transverse momentum satisfies $u_T < 15\, GeV$.

The final CDF sample consists of 470,126 $W\rightarrow e\nu$ candidates and 624,708 $W\rightarrow \mu\nu$ candidates.

D\O, in their measurement, analyzes $4.3\, fb^{-1}$ of integrated luminosity with requirements similar to the CDF event selection but uses only the electron decay channel. D\O\ selects central ($|\eta| < 1.05$) electrons with large transverse energy $E_T(e) > 25\, GeV$. The neutrino transverse energy is required to be $\slashed{E}_T(e) > 25\, GeV$ and the pair transverse mass to be in the range $50 < M_T(e,\nu) < 200\, GeV$. As in the case of CDF's event selection, highly boosted $W$ candidates are suppressed by requiring $u_T < 15\, GeV$.

The final D\O\ sample consists of 1,677,394 $W\rightarrow e\nu$ candidates. D\O\ had previously analyzed another $1\, fb^{-1}$ of data in the same channel, but acquired in lower luminosity runs of the Tevatron Collider. The higher instantaneous luminosity and corresponding higher pile-up of the $4.3\, fb^{-1}$ data acquisition period presents formidable experimental challenges to this kind of precision measurement that had to be overcome in this D\O\ analysis and will be faced by CDF in their next analysis.

\section{Calibration strategy}

The usual {\sc GEANT} based simulation of the detector response is neither fast nor precise enough to generate mass templates of the kinematical distributions to which data is compared. Both D\O\ and CDF develop dedicated Parametrized Monte Carlo Simulations (PMCS) to describe their detector response and resolution to the lepton from the $W$ boson decay. 

D\O\ and CDF calibrate the parameters in the simulation {\em in-situ} by using similar control samples, but very different strategies.

\subsection{D\O\ calibration}

The D\O\ measurement is based on a precise determination of the electron energy scale in the uranium-liquid argon (U-LAr) electromagnetic calorimeter. The central tracker is only used for direction measurement and electron identification.

The material upstream of the electromagnetic calorimeter is determined by measuring the energy fraction of the electron shower in each layer of the calorimeter. Due to the higher instantaneous luminosity of the Tevatron Collider during the data taking period of the sample analyzed by D\O, the underlying energy flow and luminosity dependence of the calorimeter gain have to be more precisely determined than in previous measurements to correctly model the response and energy deposition in each layer of the calorimeter. The high granularity of the D\O\ calorimeter is explored to measure the underlying energy flow in $W\rightarrow e\nu$ events and the dependence of the electron identification efficiency with the overall soft activity in each event. The luminosity dependence of the calorimeter response is described by a model of the ionization charge collection in the LAr gaps as a function of the luminosity.

The overall energy scale and offset are determined using $Z\rightarrow ee$ events by a two-dimensional binned maximum likelihood fit to the invariant mass $M_Z$ and $f_Z$ distributions. The observable $f_Z$ is defined as:

\begin{equation}
f_Z = \frac{\left[E(e_1) + E(e_2)\right](1-\cos\gamma)}{M_Z}
\end{equation}
where $\gamma$ is the measured angle between the electron-positron pair and $M_Z$ is their invariant mass. The energy scale and offset are determined in bins of luminosity to validate the luminosity dependence modeling of the detector response and the results are found to be statistically consistent. The D\O\ electron energy scale is known to a precision of $0.021\%$ with uncertainty dominated by the statistical power of the $Z\rightarrow ee$ sample.

Since only the $Z\rightarrow ee$ mass is used in the determination of the overall electron energy scale, the D\O\ measurement is a measurement of the ratio $M_W/M_Z$. This statement relies on the hypothesis that all the calibrations done in the somewhat more energetic $Z$ pole is valid on the $W$ pole. This hypothesis is carefully checked in each step of the calibration and, when needed, the non-linearity between the two close energy regimes is accordingly modeled. Measuring the ratio $M_W/M_Z$ is not only what allows the precise calibration to be made, since the $Z$ boson mass was measured to high precision by the LEP experiments, but also grants experimental stability against uncontrolled variations of the detector condition, since systematic variations tend to cancel in the ratio.

\subsection{CDF calibration}

The CDF measurement is based on a precise determination of the lepton momentum in the their central drift chamber (COT) immersed in a $1.4\, T$ solenoid. The interaction of the charged particles with the innermost silicon detector is modeled by a highly granular lookup table that describes ionization and radiative energy losses, multiple Coulomb scattering and Compton scattering in the tracker volume. The alignment is performed with a high-purity sample of cosmic rays muons whose trajectory is fitted to a single helix through the entire detector. Further weakly constrained modes of alignment are removed by the observed difference in the $E/p$ distributions of electrons and positrons in events that pass the $W$ boson sample selection.

The overall momentum scale is determined by binned maximum-likelihood fit to mass templates around the $J/\psi\rightarrow\mu\mu$, $\Upsilon(1S)\rightarrow\mu\mu$, and $Z\rightarrow\mu\mu$ resonances. Non-uniformities of the magnetic field are corrected by measuring the dependence of the $J/\psi$ mass with the mean polar angle. Further ionization energy losses are corrected by measuring the dependence of the momentum scale with the mean $1/p_T$ of the muons.

Using the calibrated tracker momentum scale, the peak of the $E/p$ distribution from $W\rightarrow e\nu$ and $Z\rightarrow ee$ events is fitted in bins of $E_T$ to determine electron energy scale of the calorimeter response. The amount of radiative material upstream of the COT is determined by a fit to the tail of the $E/p$ distribution. The tracker momentum scale is determined with a precision of $0.009\,\%$, dominated by uncertainties in the QED radiative corrections and magnetic field non-uniformities.

\begin{figure}[h]
\centering
\includegraphics[height=5cm]{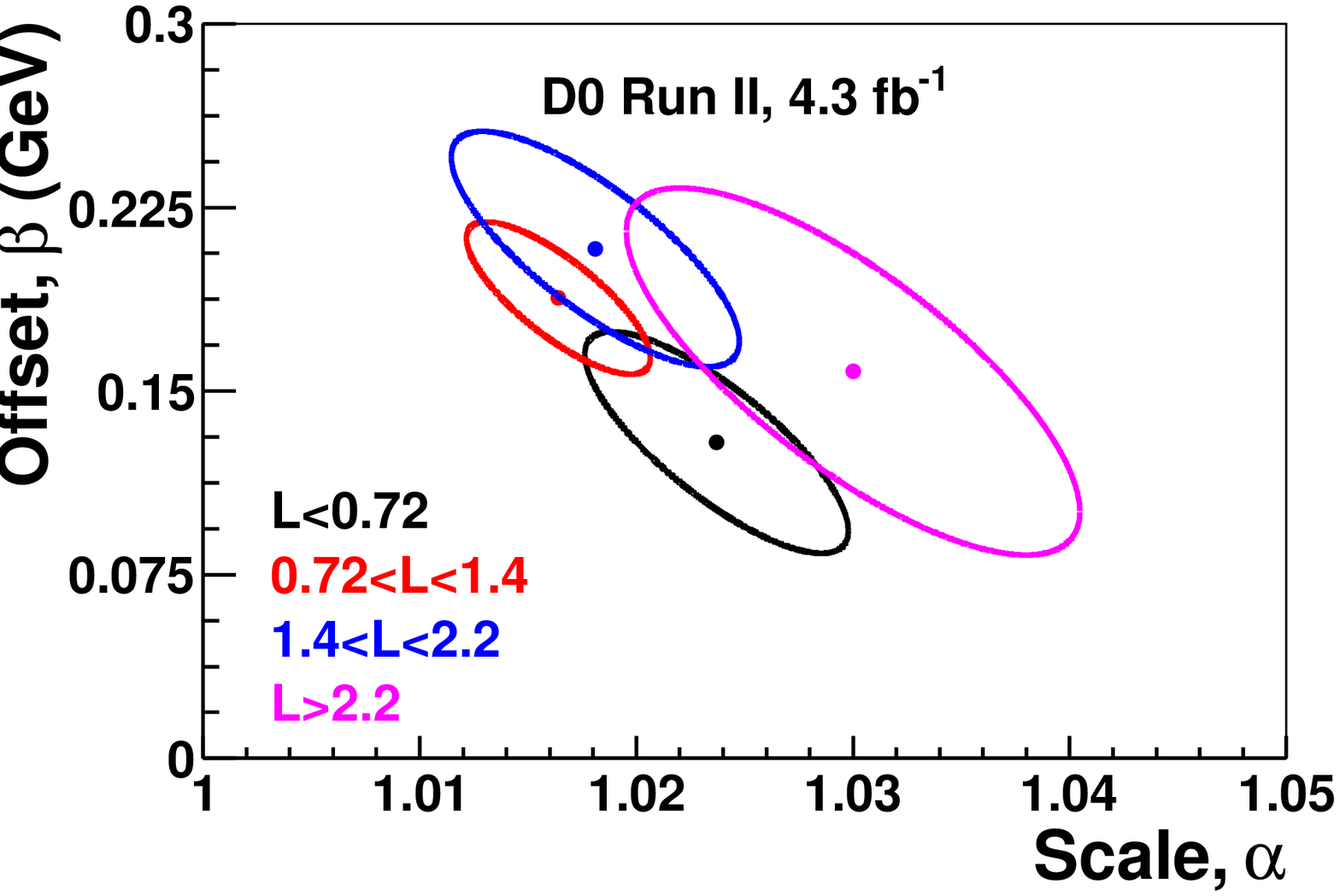}
\includegraphics[height=5.3cm]{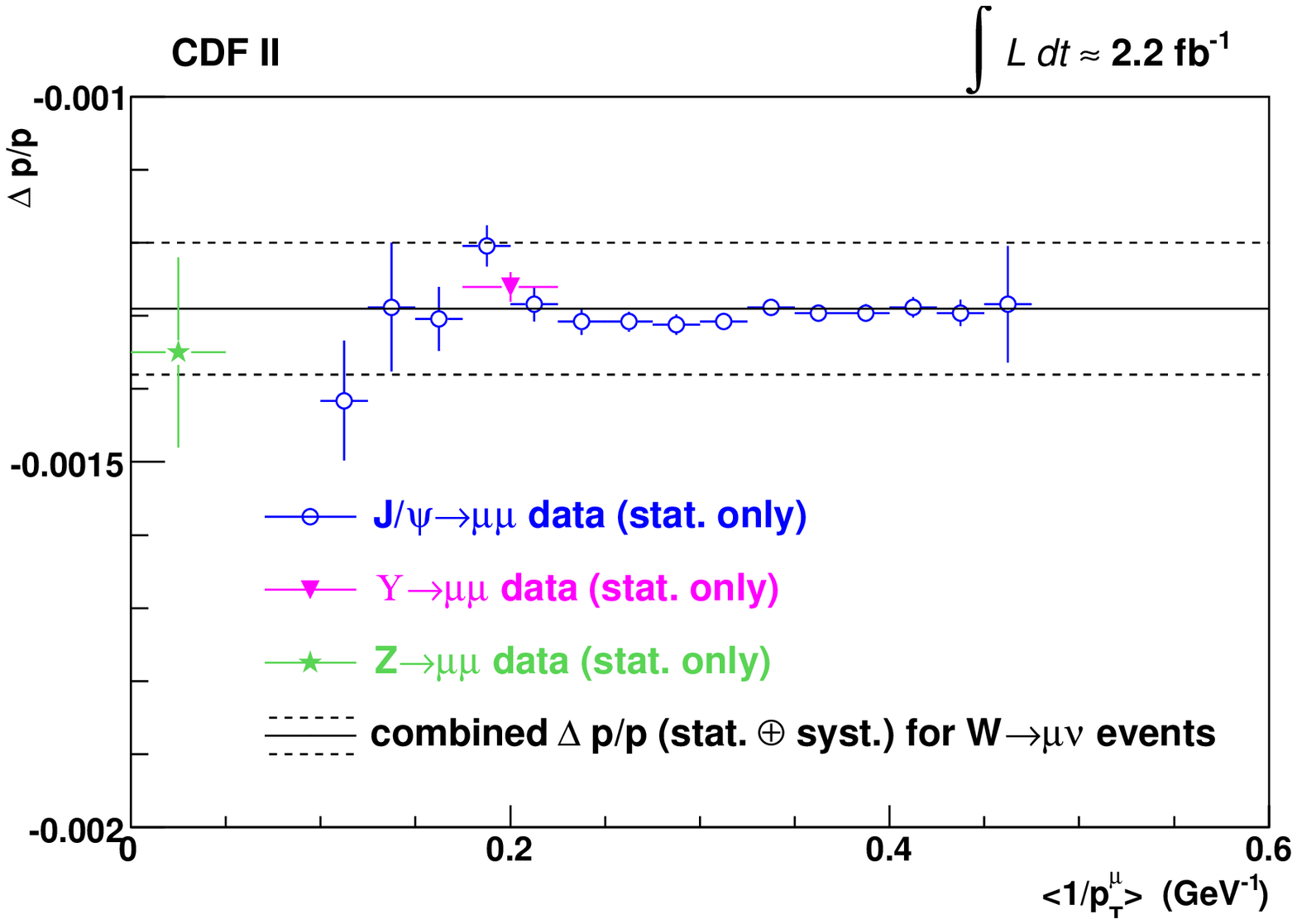}
\caption{D\O\ energy scale and offset determined in 4 different luminosity bins. CDF momentum scale determine in bins of mean $1/p_T(\mu)$ to constrain the non-linearities of the detector response.}
\end{figure}

\section{Results}

Both D\O\ and CDF perform blinded measurements. That means that throughout the analysis, a constant unknown offset is applied to the result of the fitting algorithm. The D\O\ analysis also includes an unblinded closure test where {\sc GEANT}-simulated events, with known $W$ boson mass input, are treated as data. The goal is to test the accuracy of the analysis procedure with a high statistics sample. For this measurement, a sample equivalent to $24\, fb^{-1}$ was used and closure was obtained within the statistical uncertainty of $6\, MeV$.

\subsection{D\O\ results}

After unblinding, the $W$ boson mass fit results from the D\O\ data are given in Table~\ref{t:answ}.

\begin{table}[hbtp]
\begin{center}
\caption{D\O\ and CDF results from the fits to data.  The uncertainty is only that from
    the $W$ sample statistics.  The fitting range is  $65<\mt<90\, GeV$ for transverse mass and  $32<p_T<48\, GeV$ for transverse momentum distributions .\label{t:answ}}
\vspace{0.2cm}
\begin{minipage}{0.45\textwidth}
  \begin{tabular}{c|c}
\mco{2}{c}{\textbf{D\O\ measurements}}\\
\mco{2}{c}{\ \ }\\
     Variable  &         Result (GeV)   \\  \hline\hline
      $\mt(e,\nu)$      & $\ \ \ 80.371\pm0.013\ \ \ $  \\
     $\pte$       & $      80.343\pm0.014      $ \\
     $\mete$       & $      80.355\pm0.015      $\\
  \end{tabular}
  \end{minipage}
\hspace{0.1cm}
\begin{minipage}{0.45\textwidth}
\begin{tabular}{c|c}
\mco{2}{c}{\textbf{CDF measurements}}\\
\mco{2}{c}{\ \ }\\
     Variable &         Result (GeV)        \\ \hline\hline
      $\mt(e,\nu)$   & $\ \ \ 80.408\pm0.019 \ \ \ $   \\
     $\pte$   &  $      80.393\pm0.021      $   \\
     $\mete$   &  $        80.431\pm0.025  $   \\
      $\mt(\mu,\nu)$   &  $\ \ \ 80.379\pm0.016\ \ \ $   \\
     $\ptmu$   &  $      80.348\pm0.018      $  \\
     $\metmu$   &  $      80.406\pm0.022      $   \\
  \end{tabular}
\end{minipage}
\end{center}
\end{table}


  
\begin{figure}[hbpt]
\centering
\includegraphics[width=0.495\textwidth]{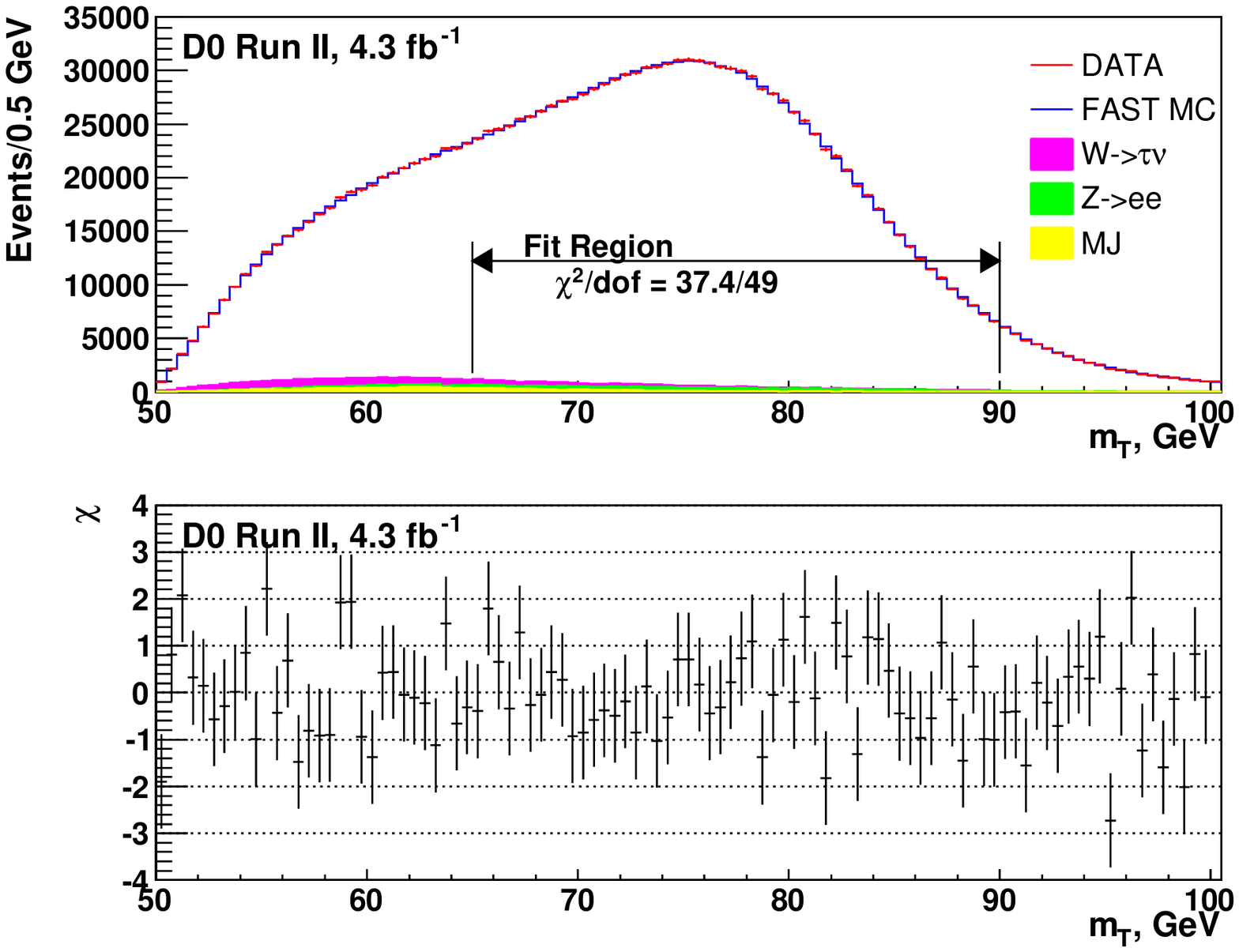}
\includegraphics[width=0.495\textwidth]{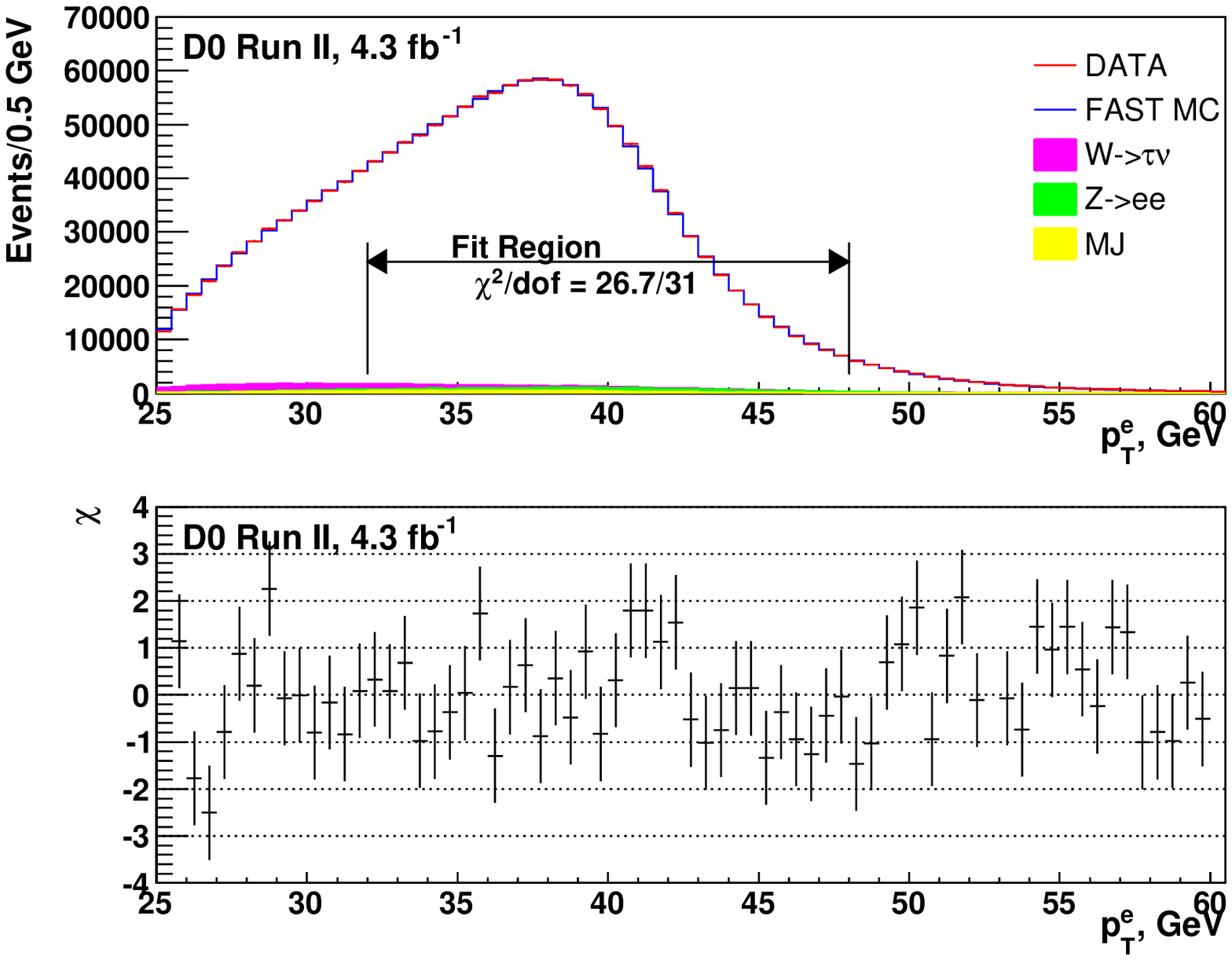}
\caption{D\O\ $M_T(e,\nu)$ and $p_T(e)$ distributions for data and PMCS simulation with backgrounds added (top) and the $\chi$ value for each bin (bottom). \label{f:d0dist}}
\end{figure}

The distributions of each variable showing the data and PMCS template with
background for the best fit value are shown in Figs~\ref{f:d0dist}.  These figures also show the bin-by-bin $\chi$ values defined as the difference between the data and template divided by the data uncertainty.

The combination of the $M_T$ and $p_T$ measurements yield a value for the $W$ boson mass of $80.367\pm 0.026\, GeV$ using only the $4.3\, fb^{-1}$ analyzed in this work. Further combining with the $1\, fb^{-1}$ previously analyzed, the new D\O\ Run II ($5.3\, fb^{-1}$) result is:

\begin{equation}
M_W(\mbox{D\O }) = 80.375 \pm 0.023\, GeV
\end{equation}

Table~\ref{t:syst2} summarizes the systematic uncertainties associated to the D\O\ measurement. Although the uncertainties are already systematically dominated, all experimental systematic uncertainties can be reduced by using a larger data sample. In the D\O\ case, a larger $Z\rightarrow ee$ sample will reduce the dominating electron energy scale uncertainty. Production uncertainties, on the other hand, are not reduced with more events and depend on further theoretical and experimental work to be better controlled.

\subsection{CDF results}

After unblinding, the $W$ boson mass fit results from the CDF data are also given in Table~\ref{t:answ}. Combining the six measurements,  the new CDF Run II ($2.2\, fb^{-1}$) result is:

\begin{equation}
M_W(\mbox{CDF}) = 80.387 \pm 0.019\, GeV
\end{equation}

Table~\ref{t:syst2} summarizes the systematic uncertainties. The CDF uncertainty is no longer dominated by lepton energy scale, but by the $W$ sample statistics and the parton distribution functions uncertainties.

%
  

\begin{table}
\centering
  \caption{Systematic uncertainties of the $M_W$ measurement. The left table shows the uncertainties for the D\O\ measurement and the right one for the CDF measurement.\label{t:syst2}}
\vspace{0.2cm}
\begin{minipage}[b]{0.43\textwidth}
\centering
  \begin{tabular}{ lccc}
          & \multicolumn{3}{c}{Unc. (MeV)} \\
   \textbf{D\O\ systematics}                          &$\mt$ & $\pte$ &  $\mete$\\
  \hline \hline
  Electron energy scale       & 16 &  17 & 16 \\
  Electron resolution         &  2 &   2 &  3 \\
  Electron shower modeling           &  4 &   6 &  7 \\
  Electron energy loss model        &  4 &   4 &  4 \\
  Hadronic recoil model             &  5 &  6 & 14 \\
  Electron efficiencies             &  1 &   3 &  5 \\
  Backgrounds                       &  2 &   2 &  2 \\ \hline
				    				     
  Parton distribution                          &  11 &  11 & 14 \\
  QED radiation                        &  7 &   7 &  9 \\
  $p_T(W)$ model                 &  2 &   5 &  2 \\ 

  \end{tabular}
\end{minipage}
\hspace{2cm}
\begin{minipage}[b]{0.43\textwidth}
\centering
  \begin{tabular}{ lc}
   
   \textbf{CDF systematics}                          & Unc. (MeV)\\
  \hline \hline

  Lepton energy scale       & \\
\ \ \ \ \ \ \ \ \ \ and resolution    & 7\\
  Recoil scale and & \\
\ \ \ \ \ \ \ \ \ \ and resolution    & 6\\
Lepton removal           &  2 \\
  Backgrounds       &  3 \\\hline
  Parton distributions             &  10 \\
  QED radiation                       &  4 \\
  $p_T(W)$ model             &  5 \\
\end{tabular}
\end{minipage}
\end{table}  

\subsection{Combination}

The two measurements described in this note were combined using the BLUE method with the older Run I and Run 0 measurements of the $W$ boson mass done by D\O\ and CDF~\cite{TevatronElectroweakWorkingGroup:2012gb}. The statistical uncertainties and systematic uncertainties, except those associated with production modeling, are taken to be uncorrelated.

Production model and theory uncertainties are partially correlated.  The minimum value of the CDF and D\O\ uncertainties for each source is assumed to be $100\%$ correlated, and the remainder for that source is assumed to be uncorrelated.   One exception is the parton distribution function uncertainty for the D\O\ measurement in Run I.  This measurements used wider eta coverage and is only $70\%$ correlated with the other measurements. In each measurement, the assumed value of the $W$ boson width is slightly different and corrected to the Standard Model predicted value of $2.0922\pm 0.0015\, GeV$ in the running-width scheme using the newly obtained $W$ boson mass world average. After all corrections, the new Tevatron combination for the value of the $W$ boson mass is:

\begin{equation}
M_W(\mbox{Tevatron}) =80.387\pm 0.016\, GeV
\end{equation}

Further combining this result with the LEP direct measurements, which are considered to be completely uncorrelated with the Tevatron result, the new world average value of the $W$ boson mass is:

\begin{equation}
M_W(\mbox{WA}) =80.385\pm 0.015\, GeV
\end{equation}

The $\chi^2$ of the combination is 4.3 for 7 degrees of freedom with a probability of $74\%$. The results is strongly dominated by the D\O\ and CDF Run II measurements.

\section{Model and theoretical uncertainties}

In the D\O , but even more so in the CDF measurement, the uncertainty in the $W$ boson mass is dominated by model and theoretical uncertainties. In particular, the parton distribution function (PDF) uncertainty is already the most important uncertainty in the CDF measurement and will be in the next D\O\ measurement. To further improve the precision of the measurements, these uncertainties have to be controlled by improving both experimental techniques and theoretical understanding of the processes involved.

The PDF uncertainties are, to a large extent, an acceptance uncertainty that are introduced by the lepton acceptance requirement made by both D\O\ and CDF. In Run I, D\O\ extended the $\eta$ coverage of the $W$ sample in the $W$ boson mass measurement~\cite{OldW}. The forward region brings other experimental challenges, such as the larger amount of underlying energy flowing through the detector, but the wide coverage of the D\O\ calorimeter must be explored in the near future. The relevant $u$ and $d$ quarks PDF can also be constrained at high mass scales by measuring the $W$ charge asymmetry at the Tevatron and introducing the result in global QCD fits. Improved calculations of $W$ production and decay in hadron colliders can also be used to reduce uncertainties associated to higher order QED and QCD corrections~\cite{NewQED}. Finally, recently proposed kinematical distributions that carry more mass information than the transverse mass can be attempted to extract the $W$ boson mass with less sensitivity to the systematic uncertainties~\cite{Rujula:2011qn}.

\section{Higgs constraints from global electroweak fit}

The updated $W$ boson mass world average can be used together with the electroweak precision measurements performed at LEP, Tevatron and SLC~\cite{ALEPH:2005ab} to indirectly measure the Standard Model Higgs boson mass. The value, prior to the two measurements described in this note was $M_H = 92^{+34}_{-26}\, GeV$. With the Tevatron $W$ boson mass measurements presented here, the new \textit{Tevatron Electroweak Working Group} indirect value of the Higgs boson mass is~\cite{TevatronElectroweakWorkingGroup:2012gb}:

\begin{equation}
M_H(\mbox{indirect}) = 94^{+29}_{-24}\, GeV
\end{equation}

Using the full $10\, fb^{-1}$ recorded by both D\O\ and CDF, the Tevatron experiments can reduce the uncertainty in the $W$ boson mass to $10\, MeV$. Such precision, together with the planned improvements on the top quark mass measurement, will allow a confrontation between the indirect and potential direct measurement of the Higgs boson mass with similar precisions. Even after the Higgs boson mass has been measured to high precision, the $W$ boson mass will continue to be the most important parameter in the determination of the global consistency of the electroweak sector of the Standard Model.

\section{Conclusions}

\begin{figure}
\centering
\begin{minipage}{0.45\textwidth}
\centering
\includegraphics[width=\textwidth]{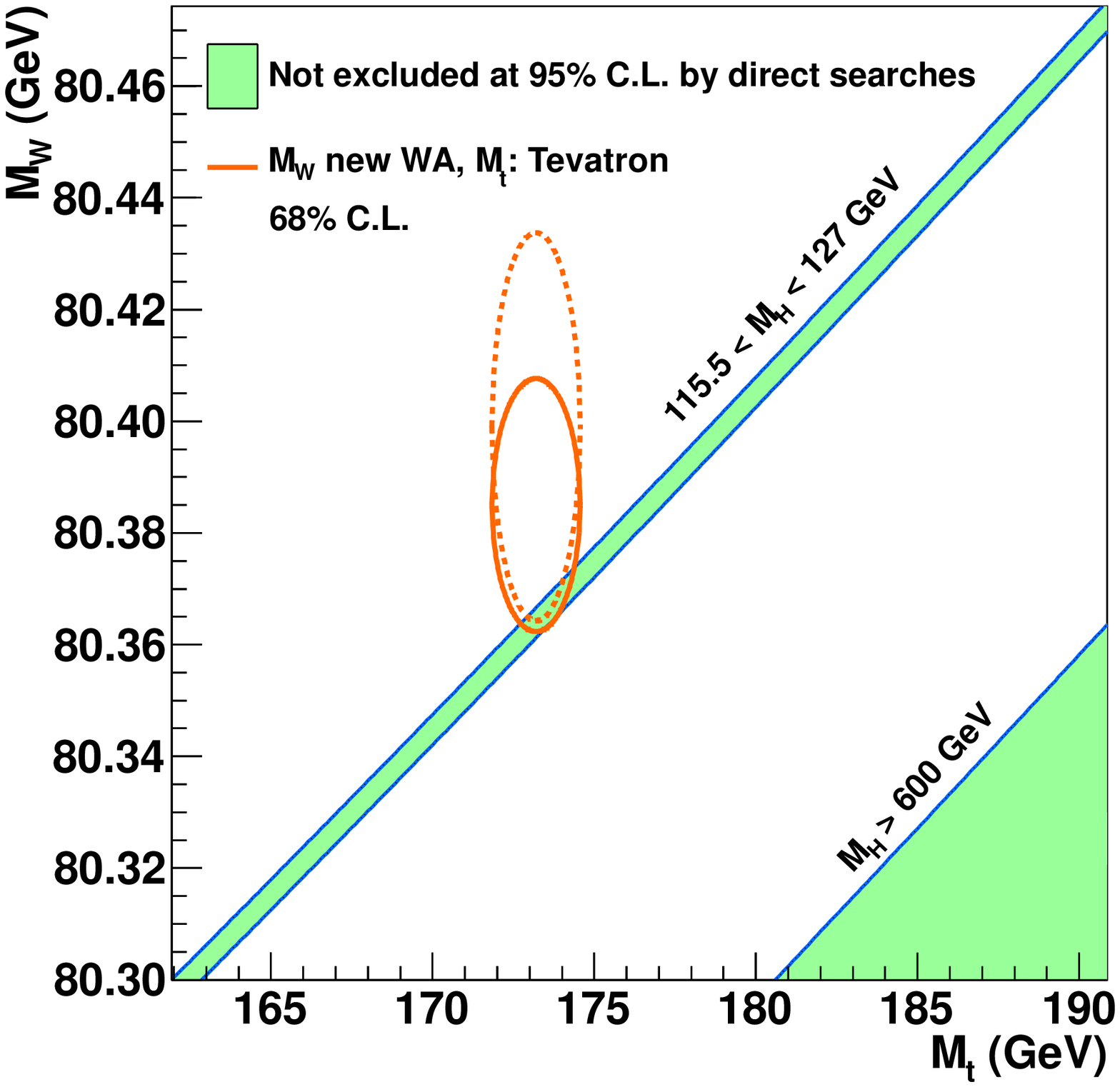}
\end{minipage}
\begin{minipage}{0.45\textwidth}
\centering
\includegraphics[width=\textwidth]{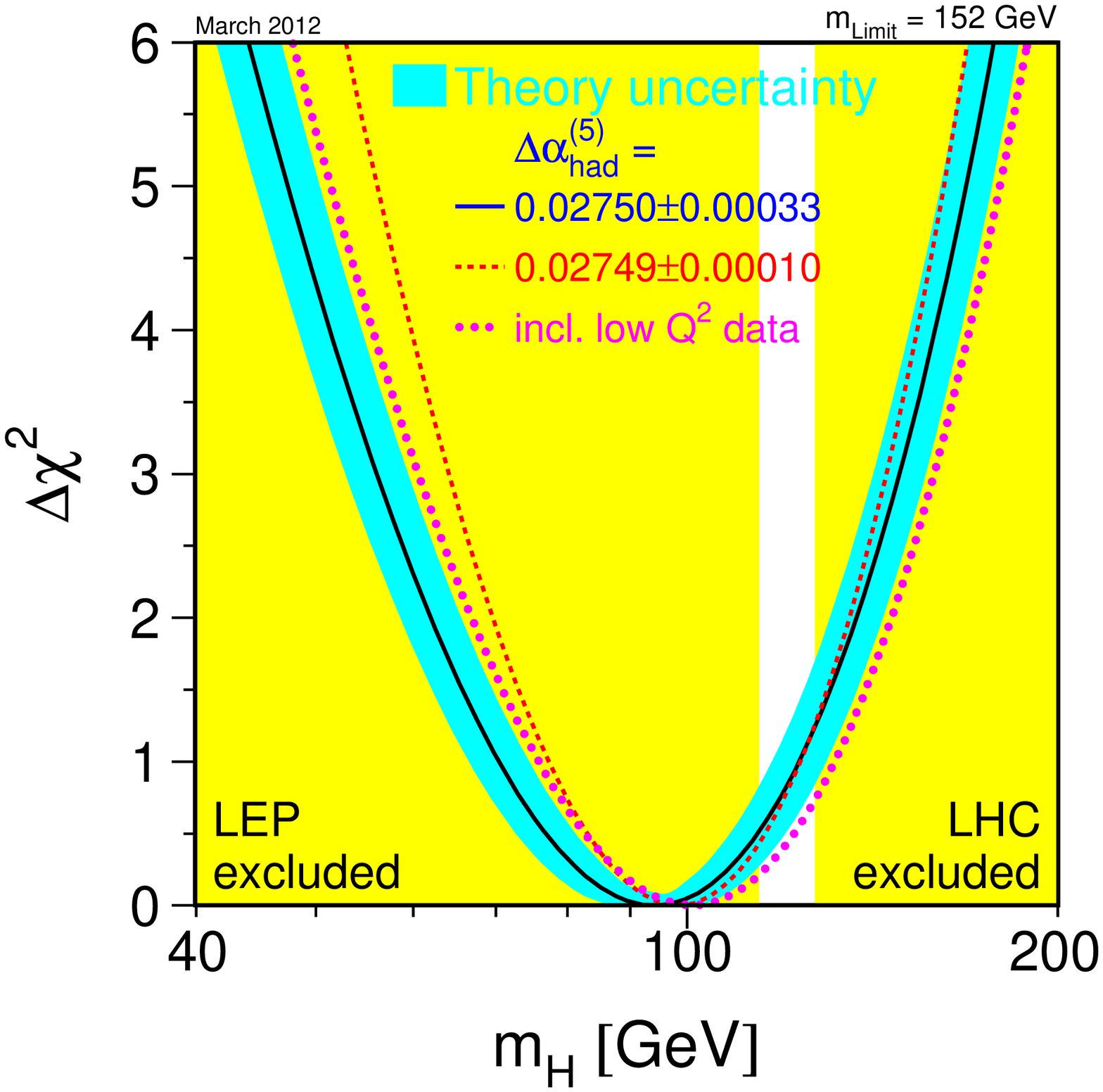}
\end{minipage}
\caption{On the left, comparison of the Standard Model prediction of the $W$ boson mass for varying values of the Higgs boson mass compared to the direct measurement. The previous world average, prior to the measurements presented in this note, is marked by the dashed ellipse. The green band is a purely experimental exclusion and does not include limits from perturbative unitarity. On the right, the $\chi^2$ of the updated global electroweak fit prediction of the Higgs boson mass.\label{f:mtmw}}
\end{figure}

The $W$ boson mass was measured by both D\O\ and CDF collaborations with precision at least as good as the world average average prior to these measurements~\cite{Abazov:2012bv}$^,$~\cite{Aaltonen:2012bp}. Despite using very different calibration procedures, all D\O\ and CDF measurements are consistent. The new $W$ boson mass world average is consistent with the Standard Model prediction for a low mass Higgs boson and strongly disfavors a high mass Higgs, as can be seen in Fig.~\ref{f:mtmw}.

\section*{Acknowledgments}
I would like to thank the organizers of the Moriond QCD conference for the opportunity to give the talk in such nice scientific environment. I would also like to acknowledge the D\O\ and CDF collaborations, the Tevatron Accelerator Division and the corresponding funding agencies, for the outstanding work for more than 20 years that allowed works like these ones to be produced.

\end{document}